\documentclass[prd,twocolumn,superscriptaddress,longbibliography,preprintnumbers,floatfix,nofootinbib]{revtex4-2}
\usepackage{url}
\usepackage{xspace}
\usepackage{dsfont}
\usepackage{amssymb}
\usepackage{amsmath}
\usepackage{graphicx}
\usepackage[caption=false]{subfig}
\usepackage[colorlinks=true,citecolor=blue]{hyperref}
\usepackage{multirow}
\usepackage{hhline}
\usepackage{xcolor}
\usepackage{comment}
\usepackage{cleveref}
\usepackage[acronym]{glossaries}
\newacronym{NLO}{NLO}{next-to-leading order}
\newacronym{QCD}{QCD}{quantum chromodynamics}
\newacronym{NSBI}{NSBI}{neural simulation-based inference}

\graphicspath {{figures/}}

\begin{document}

%\preprint{APS/123-QED}
\title{Stay Positive: Neural Refinement of Sample Weights}
% \title{Staying Positive with Neural Refinement of Sample Weights}

\author{Benjamin Nachman}%
\affiliation{%
  Physics Division, Lawrence Berkeley National Laboratory, Berkeley, CA 94720, USA
}%
\affiliation{%
  Fundamental Physics Directorate, SLAC National Accelerator Laboratory, Menlo Park, CA 94025, USA
}%
\affiliation{%
  Department of Particle Physics and Astrophysics, Stanford University, Stanford, CA 94305, USA
}%
\author{Dennis Noll}%
\email{dnoll@lbl.gov}
\affiliation{%
  Physics Division, Lawrence Berkeley National Laboratory, Berkeley, CA 94720, USA
}%

\date{\today}
\begin{abstract}
Monte Carlo simulations are an essential tool in particle physics data analysis.
Events are typically generated alongside weights that redistribute the cross section of the simulated process across the phase space.
These weights can be negative, and several post-hoc methods have been developed to eliminate or mitigate the negative values.
All of these methods share the common strategy of approximating the average weight as a function of phase space.
We introduce an alternative approach, which, instead of reweighting to the average, refines the initial weights with a scaling transformation, utilizing a phase space-dependent factor.
Since this new refinement method does not need to model the full weight distribution, it can be more accurate.
High-dimensional and unbinned phase space is processed using neural networks for the refinement method.
In addition to the refinement method, we introduce a new resampling protocol, which can be used in conjunction with any weight transformation to not only preserve the average weight but also the statistical uncertainties of the initial distribution.
Using both realistic and synthetic examples, we show that the new neural refinement method is able to match or exceed the accuracy of similar weight transformations and that the new resampling protocol is simpler in implementation than previous methods while exhibiting equivalent statistical properties.
\end{abstract}

\maketitle

%\tableofcontents

\section{Introduction}
Statistical analysis in particle physics usually involves the comparison of simulated and recorded data.
The simulated data are generated with Monte Carlo methods, and the events are often weighted due to the complex nature of the underlying physics.
The weights themselves are not physical and only the sum of all weights at a given phase-space point corresponds to an observable.
The spread of weights reduces the statistical power of a synthetic dataset.
Sometimes, weights can be negative.
As cross sections are non-negative, the existence of negative weights implies a spread in weights and thus a dilution in statistical power.
Negative weights may also lead to other complications in downstream analyses.
There are multiple origins of negative weights, including higher-order perturbative corrections and background subtraction.

Several methods have been proposed to reduce the contribution from negative weights.
For example, multiple algorithm-specific modifications to higher-order calculations have been studied for reducing negative weight contributions~\cite{Danziger:2021xvr,Alioli:2010xd,Frederix:2020trv}.
Complementary approaches are algorithm-agnostic techniques that reduce negative weights post-hoc~\cite{Andersen:2020sjs,Nachman:2020fff,Andersen:2021mvw,Andersen:2023cku,Andersen:2024mqh}.
These tools can also have desirable statistical properties like uncertainty conservation~\cite{Nachman:2020fff}.
All of these methods have a similar strategy - the weight $w_i$ for event $i$ is replaced by the average weight for the corresponding point in phase space $x_i$.
A key challenge with these approaches is that this reassignment of weights can be highly non-trivial, especially if the conditional probability density $p(w|x)$ has a large variance.

We make a simple, but powerful observation that instead of reweighting events by replacing $w_i$ with the average, we can refine the $w_i$ of each event to remove negative weights.
An extreme example, which we will discuss in more detail below, is the case where there is a non-trivial $p(w|x)$, yet all weights are non-negative.
Reweighting methods still act non-trivially on such events, while our new strategy can leave them unchanged.
This weight refining has several practical advantages over reweighting that we will highlight with numerical examples.
We use neural networks in all our examples since they scale well to many dimensions, but the refinement strategy can be adapted to any local estimation of weights.

In addition to the new refinement method, we introduce a novel resampling protocol that restores the statistical uncertainties of the initial distribution after the weight transformation.
Our new protocol is easier to implement than existing methods~\cite{Nachman:2020fff} while exhibiting equivalent statistical properties.
It is compatible with both reweighting and refinement, as well as histogram-based or comparable weight transformations.

This paper is organized as follows.
\Cref{sec:methods} introduces the formal properties of reweighting and refinement, and numerical examples are provided in \cref{sec:results}.
\Cref{sec:resampling} introduces the new resampling protocol and demonstrates its application on a synthetic example.
The paper ends with conclusions and an outlook in \cref{sec:conclusions}.

\section{Neural Refinement}
\label{sec:methods}

First, we briefly review neural reweighting before introducing weight refinement.
The goal of reweighting is to replace weights $w_i$ by $\int\text{d}x\,p(z|x)w(x,z)$, where the latent variable $z$ is the source of randomness for the weights at a fixed phase space point $x$.
In other words, reweighting replaces individual weights by the average weight for a given phase space point $x=x_i$.
This can be accomplished with the following loss functional:

\begin{align}\nonumber
  \mathcal{L}[g]=&-\int \text{d}x\text{d}z \,p(x,z)w(x,z)\log(g(x))\\\label{eq:bce}&\hspace{2mm}-\int \text{d}x \,p(x)\log(1-g(x))\,,
\end{align}
where we are working in the continuum limit, where a sum over samples is replaced with an integral weighted by the probability density, as this simplifies the functional analysis.
\Cref{eq:bce} is the standard binary cross-entropy, but with the same samples in both terms.
The only difference between the first and second term is that the second term has the weights.
The functional optimum of \cref{eq:bce} is

\begin{align}
\label{eq:reweight}
  w_i'\equiv\frac{g_\text{reweight}^*(x_i)}{1-g_\text{reweight}^*(x_i)}=\int \text{d}z\,p(z|x_i) w(x_i,z)\,.
\end{align}

In practice, $g$ is parameterized as a neural network and optimized using gradient-based algorithms.
The main feature of \cref{eq:reweight} that we seek to address is that the weights $w_i$ and $w_i'$ are often quite different, even when all of the weights are non-negative.
Instead of collapsing all weights to a single value, we investigate a different scheme in which the positive and negative weights are separately shifted in such a way that nothing happens in the case where all weights are non-negative.

The new method refines the original sample weights through a scaling transformation, utilizing a phase space-dependent factor, called the refinement factor.
The samples that originally had a positive (negative) weight are denoted as positive (negative) samples.
The absolute values of the weights of negative samples are denoted as pseudo-positive weights.
The negative samples, which are assigned their pseudo-positive weight, are denoted as pseudo-positive samples.
The refinement factor is derived from the likelihood ratio between the distribution of pseudo-positive samples and the distribution of positive samples.

Each phase space bin has its own refinement factor, and the refined weights are calculated as follows:
For positive samples, the refined weight is given by the refinement factor multiplied by the original weight.
For negative samples, the refined weight is given by the refinement factor multiplied by the pseudo-positive weight.

The likelihood ratio between positive and pseudo-positive samples, and thus the refinement factor, is learned in a phase-space-dependent way.
By utilizing a neural network to model the likelihood ratio, as with neural reweighting, this method is completely unbinned and capable of handling high-dimensional and even variable-dimensional data.
The refinement factor is evaluated and applied on a sample-by-sample basis.

The loss functional for the new refining method in the continuum limit is as follows:

\begin{align}
    \mathcal{L}[g] =
    &-\int \text{d}x\text{d}z\, p(x,z) \frac{|w(x,z)| + w(x,z)}{2} \log(g(x))\nonumber\\
    &-\int \text{d}x\text{d}z \,p(x,z) \frac{|w(x,z)| - w(x,z)}{2} \log(1-g(x))\,,
    \label{eq:loss}
\end{align}
where the first term corresponds to the positive samples and the second term corresponds to the pseudo-positive samples.
The function $g^*_\text{refine}$ that optimizes \cref{eq:loss} results in

\begin{align}
    \label{eq:refinefactor}
    \frac{g^*_\text{refine}(x)}{1-g^*_\text{refine}(x)}
    = \frac{\int \text{d}z\, p(z|x) (|w|+w)}{\int \text{d}z\, p(z|x) (|w|-w)}\,.
\end{align}

And with the definition of
\begin{align}
    w_{+} &= \int \text{d}z\, p(z|x) \frac{|w|+w}{2} \quad\text{and}\\
    |w_{-}| &= \int \text{d}z\, p(z|x) \frac{|w|-w}{2} \quad\text{and}\\
    r &=\frac{|w_{-}|}{w_{+}} = \frac{1-g^*_\text{refine}(x)}{g^*_\text{refine}(x)}\,,
\end{align}
we apply the following transformation:
\begin{align}
  \label{eq:refine}
  w_i\mapsto \tilde{w}_i\equiv |w_i|\,\frac{1-r(x_i)}{1+r(x_i)}\,.
\end{align}

Like reweighting, refinement also reproduces the average weight:
\begin{align}\nonumber
    \int\text{d}z\, & p(z|x) \tilde{w}(z,x)
    =\frac{1 - r(x)}{1 + r(x)} \int\text{d}z\, p(z|x) |w(x,z)|\\\nonumber
    &=\frac{1 - |w_{-}|/w_{+}}{1 + |w_{-}|/w_{+}} \int \text{d}z \,p(z|x) |w(x,z)|\\\nonumber
    &=\frac{w_{+} - |w_{-}|}{w_{+} + |w_{-}|} \int \text{d}z \,p(z|x) |w(x,z)|\\\nonumber
    &=\frac{\int \text{d}z \,p(z|x) w(x,z)}{\int \text{d}z \,p(z|x) |w(x,z)|}\int \text{d}z \,p(z|x) |w(x,z)|\\
    &=\int \text{d}z \,p(z|x) w(x,z)\,.
\end{align}
A critical difference to reweighting is that $\tilde{w}_i=w_i$ when all weights are non-negative.

\begin{figure}
  \includegraphics[width=\linewidth]{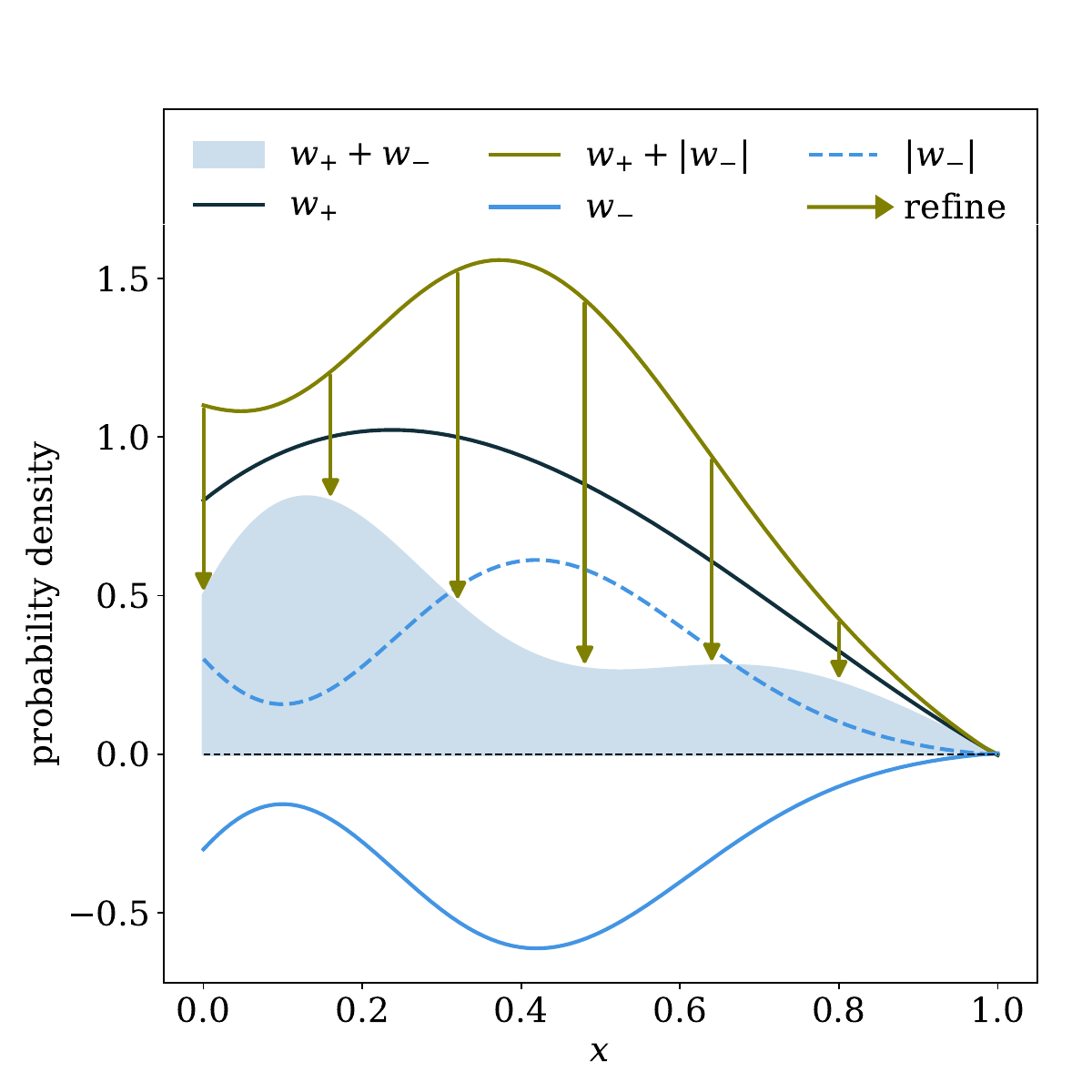}
  \caption{\label{fig:schema} Graphical illustration of the refinement method and the distribution of samples with different weights. }
\end{figure}

\Cref{fig:schema} provides a graphical illustration of the refinement method and displays the relationships between the distributions of samples with different weights.
The figure shows the distributions of positive and negative samples ($w_{+} + w_{-}$), positive samples ($w_{+}$), negative samples ($w_{-}$), pseudo-positive samples ($|w_{-}|$), and the combined distribution of positive and pseudo-positive samples ($w_{+} + |w_{-}|$).
The refinement process is visually represented as a scaling transformation from the distribution of positive and pseudo-positive samples to the distribution of positive and negative samples.
The refinement factor, depicted as arrows, can be computed from the likelihood ratio between the distributions of pseudo-positive and positive samples, as given by \cref{eq:refinefactor}.

After the refinement, the local resampling strategy introduced in Ref.~\cite{reweighter} can be used to restore statistical properties and uncertainties of the samples.

The refinement method offers several potential benefits over reweighting, including:
\begin{itemize}
  \item \textbf{Simplified task}: The network is only required to learn the refinement between the distributions of pseudo-positive and positive weights. This means that the network does not need to learn the complete normalization and shape of the distribution of positive and negative weights, which can be a more complex task.
  \item \textbf{Preservation of weight spectrum}: As the refinement factor is a multiplicative factor within a particular phase space bin, the refinement method preserves the relative spectrum of weights among the positive and pseudo-positive weights, respectively, in each phase space bin.
  \item \textbf{Improved extrapolation}: The refinement method can exhibit better extrapolation to regions where the distribution of weights is non-trivial, but the ratio between pseudo-positive and positive weights is well-behaved and learnable by a neural network.
  \item \textbf{Robustness to negative density}: The refinement method is resilient to the occurrence of samples from a phase space with a negative density, whereas similar approaches are not and show diverging behavior in these cases. While a negative density is unphysical per se, it can occur in several cases as described below.
\end{itemize}
Each of these benefits is individually demonstrated in the case studies presented in the next section, providing a comprehensive evaluation of the advantages of the refinement method.

\section{Case Studies}
\label{sec:results}
Several case studies are performed to showcase the working principle of the weight refinement and to illustrate the successful application of the method.
As a benchmark, we compare with reweighting, implementing the neural resampling method from \cite{reweighter}.

In each case study, 80\% of the samples are used for training and 20\% of the samples are used for testing.
The figures involving the evaluation of trained networks show the distributions of samples used for testing only.

All neural networks feature a feed-forward architecture, consisting of two hidden layers with 128 nodes each, and utilize the ReLU activation function.
If not noted differently, the networks are trained for 10 epochs with a batch size of 1024 samples using the Adam optimizer with a learning rate decaying exponentially from $10^{-3}$ to $10^{-6}$ during the complete training process.
As the reweighter method uses each sample twice in each epoch, once with its original weight and once with a nominal weight, the neural network for the reweighter method is effectively trained on twice the number of samples per epoch~\cite{reweighter}.
The hyperparameters were not extensively optimized, but small changes in the setup did not result in qualitative changes in performance.

All neural networks used in the case studies were implemented and trained using the TensorFlow~\cite{tensorflow2015-whitepaper} and Keras~\cite{chollet2015keras} deep learning frameworks.

The first (last) bins of all shown histograms serve as underflow (overflow) bins, and include all samples with values below (above) the plotted range.
The scatter plots show 1000 randomly selected samples from the test set for each shown distribution.

\subsection{Top Quark Pair Production}
The first case study shows the performance of the refiner and the reweighter method on a realistic physics dataset with top quark pair production at next-to-leading order (NLO) in quantum chromodynamics (QCD).
It is based on a total number of 2M events, which were generated as described in Ref.~\cite{reweighter} and briefly summarized in the following.
Fixed-order matrix elements are generated using MG5\_aMC@NLO 5.2.7.2~\cite{Alwall:2014hca} interfaced with the NNPDF 2.3 NLO parton density function set~\cite{Ball:2012cx}.
These calculations use \textsc{MadLoop}~2.7.2~\cite{Hirschi:2011pa,Alwall:2014hca}, \textsc{Ninja}~1.2.0~\cite{Mastrolia:2012bu,Peraro:2014cba}, \textsc{CutTools}~1.9.3~\cite{Ossola:2007ax}, and \textsc{OneLoop}~3.6 ~\cite{vanHameren:2010cp,vanHameren:2009dr}.
Both top quarks decay leptonically via $t\bar{t}\rightarrow b\bar{b}\mu^+\mu^-\nu\bar{\nu}$.
The default FKS subtraction scheme~\cite{Frixione:1995ms,Frixione:1997np}, is used to match the resulting samples with the \textsc{Pythia} 8.230~\cite{Sjostrand:2014zea,Sjostrand:2006za} parton shower, with its default settings.
The output of \textsc{Pythia} is recorded in the \textsc{HepMC}~2.06.09~\cite{Dobbs:684090} format and then processed with \textsc{Delphes}~3.4.2~\cite{deFavereau:2013fsa}.
The reconstructed particles are then clustered into $R = 0.4$ anti-$k_t$ jets~\cite{Cacciari:2011ma} with \textsc{FastJet}~\cite{Cacciari:2005hq}.
Jets are $b$-tagged using the flavor tagging module in \textsc{Delphes}.

The dataset consists of events with two charged leptons, two neutrinos, two $b$-jets, and a variable number of additional jets, up to a maximum of 15 objects per event.
Each object is characterized by a $p_T$, $\eta$, $\phi$, mass, and identification (integer encoding type).
Due to the variable number of jets per event, the weight transformation task on this dataset is inherently variable-dimensional.
For this example, we simplify the problem by zero padding and then using a fully connected neural network.
Architectures naturally able to ingest variable-length inputs may be more efficient, although that was not a challenge in this study.
To assess the performance of the methods, 10 independent neural networks with different random initializations are trained for both the reweighting and refinement tasks.

\begin{figure*}
  \centering
  \subfloat[]{\label{fig:tt_ensemble:raw}
    \centering
    \includegraphics[width=0.32\linewidth]{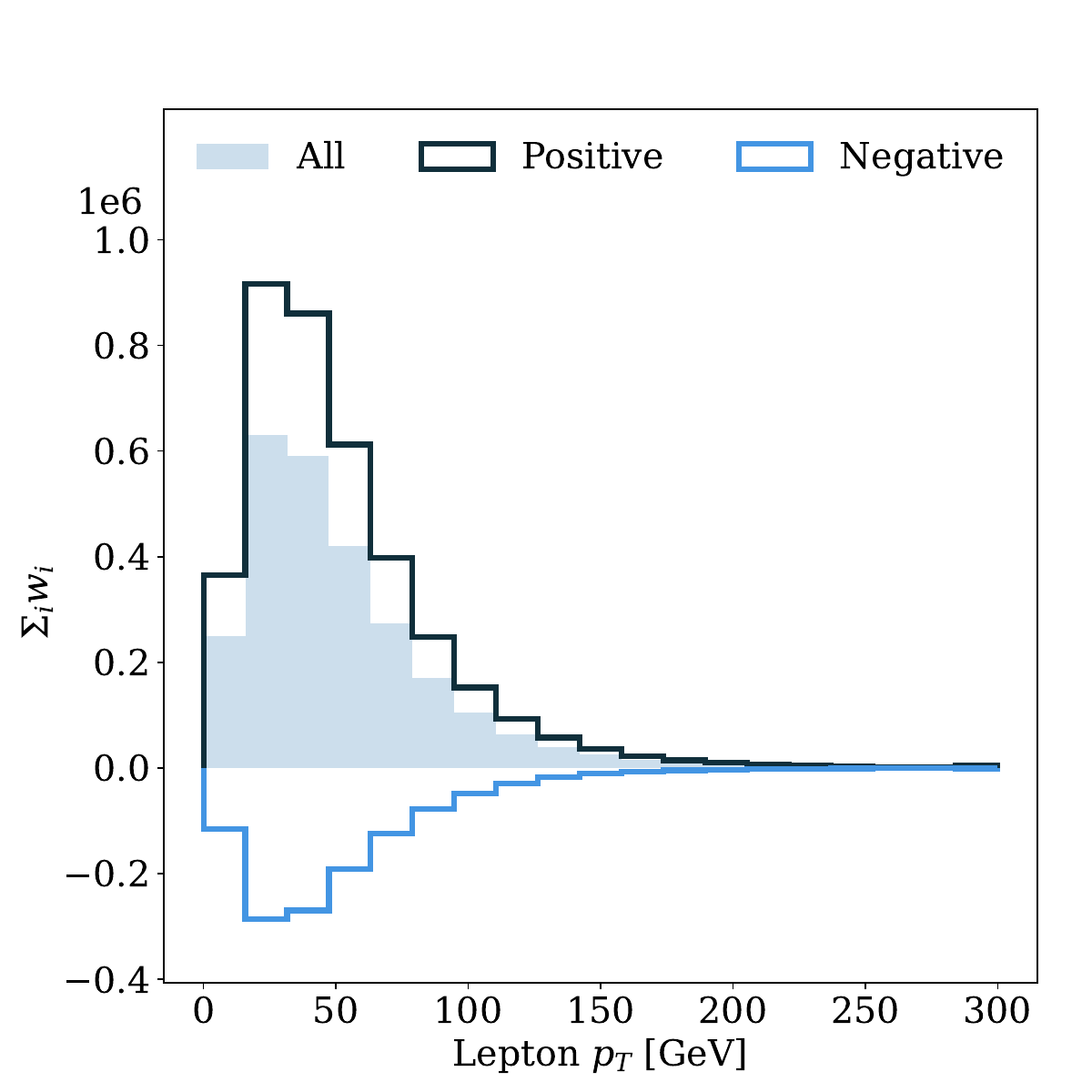}
  }%no space
  \hfill
  \subfloat[]{\label{fig:tt_ensemble:counts}
    \centering
    \includegraphics[width=0.32\linewidth]{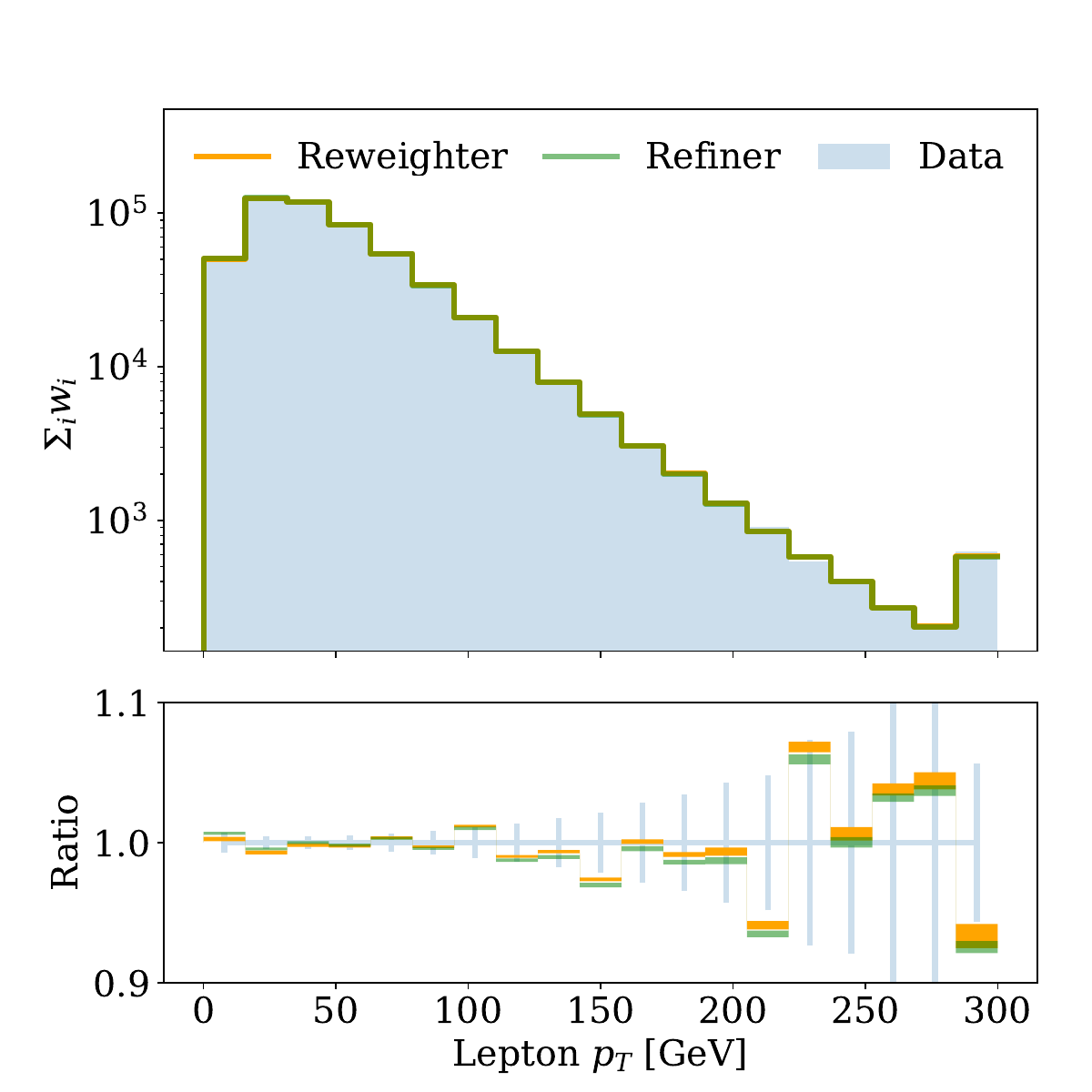}
  }%no space
  \hfill
  \subfloat[]{\label{fig:tt_ensemble:weights}
    \centering
    \includegraphics[width=0.32\linewidth]{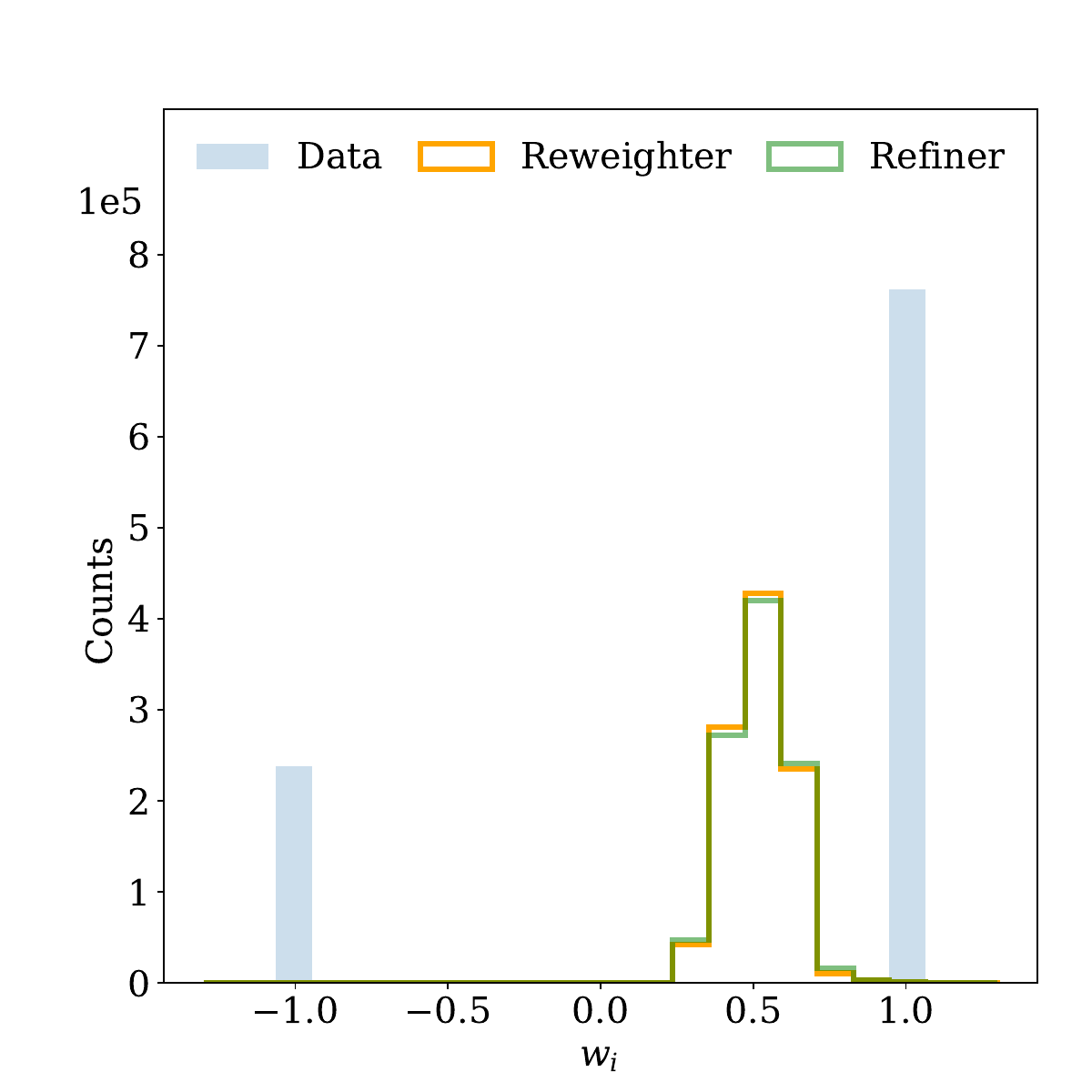}
  }
  \caption{
    Top quark pair production.
    Left: a histogram of the positively charged lepton transverse momentum broken down into positive and negative events.
    Middle: the reweighted/refined spectrum compared to the initial data (with negative weights).
    Right: the distribution of event weights before and after reweighting/refining. The error bands in the middle plot represent the spread across the 10 networks, and the vertical error bars represent the statistical uncertainty.
  }
  \label{fig:tt_ensemble}
\end{figure*}

To illustrate the impact of negative weight mitigation in the $t\bar{t}$ example, we examine the charged particle $p_T$ spectrum in \Cref{fig:tt_ensemble} (the results are similar for other observables).
In this example, about 20\% of events have negative weights.
\Cref{fig:tt_ensemble:counts} demonstrates that both reweighting and refining can match the original spectrum with negative weights (labeled `Data').
The error bands in the ratio show that both methods are robust to the network initialization, with a spread that is typically smaller than the statistical uncertainty.
There is an interplay between this spread and the hyperparameter optimization, and it would be interesting to explore this in further detail in the future.

The resulting distribution of weights is presented in \Cref{fig:tt_ensemble:weights}.
As desired, the reweighted/refined weights are non-negative and have nearly the same distribution since they have the same target.
This top quark example shows that on a relevant particle physics problem with a relatively simple weight distribution, both reweighting and refining work well.
The next sections focus on synthetic examples that highlight potential advantages of refining over reweighting.
While the trends are exaggerated, the accuracy requirements at current and future experiments may be sensitive to scaled-down versions of these setups.

\subsection{Synthetic Shape Example}

\begin{figure*}
  \centering
  \subfloat[]{\label{fig:weight_shape:raw}
  \centering
  \includegraphics[width=0.32\linewidth]{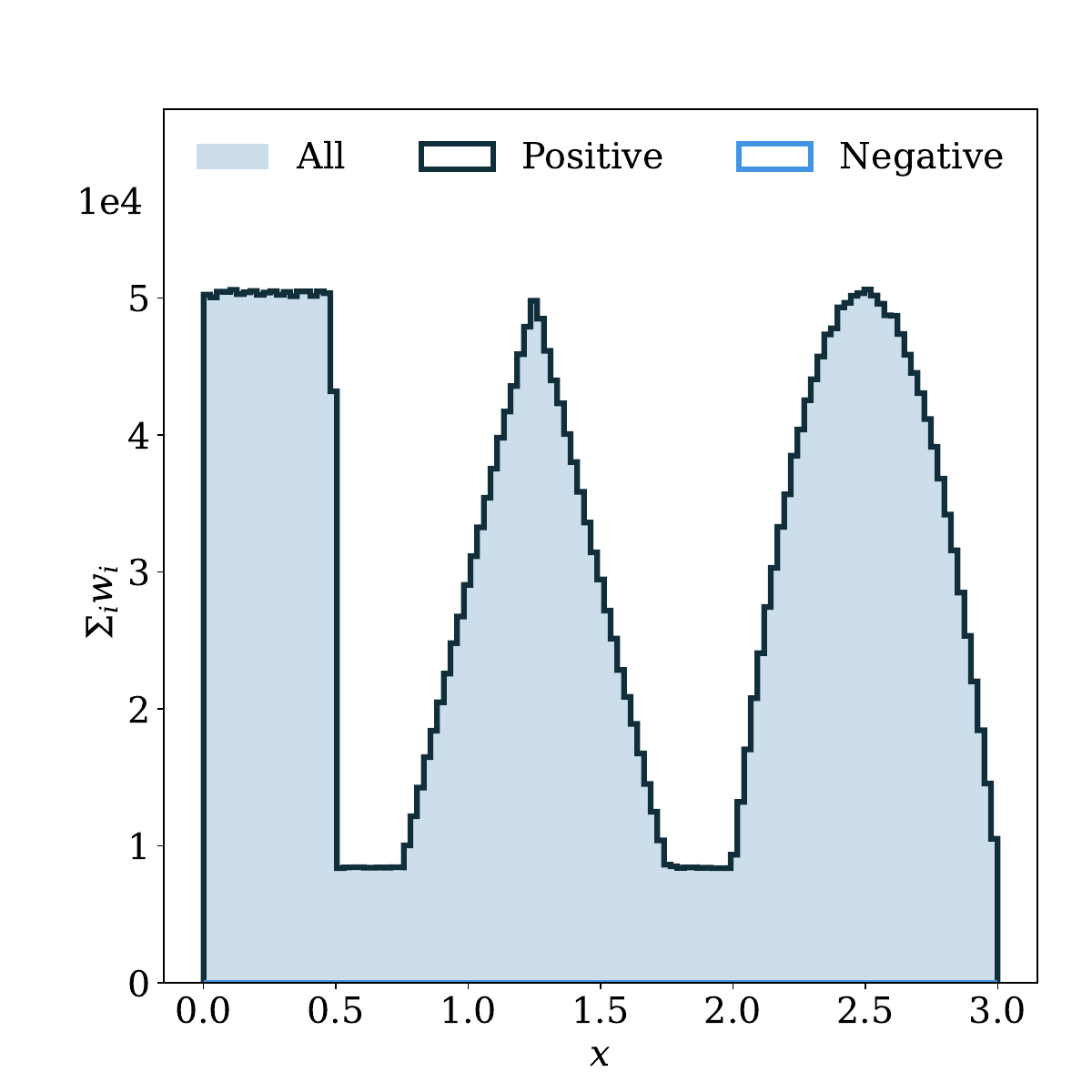}
  }%no space
  \hfill
  \subfloat[]{\label{fig:weight_shape:counts}
    \centering
    \includegraphics[width=0.32\linewidth]{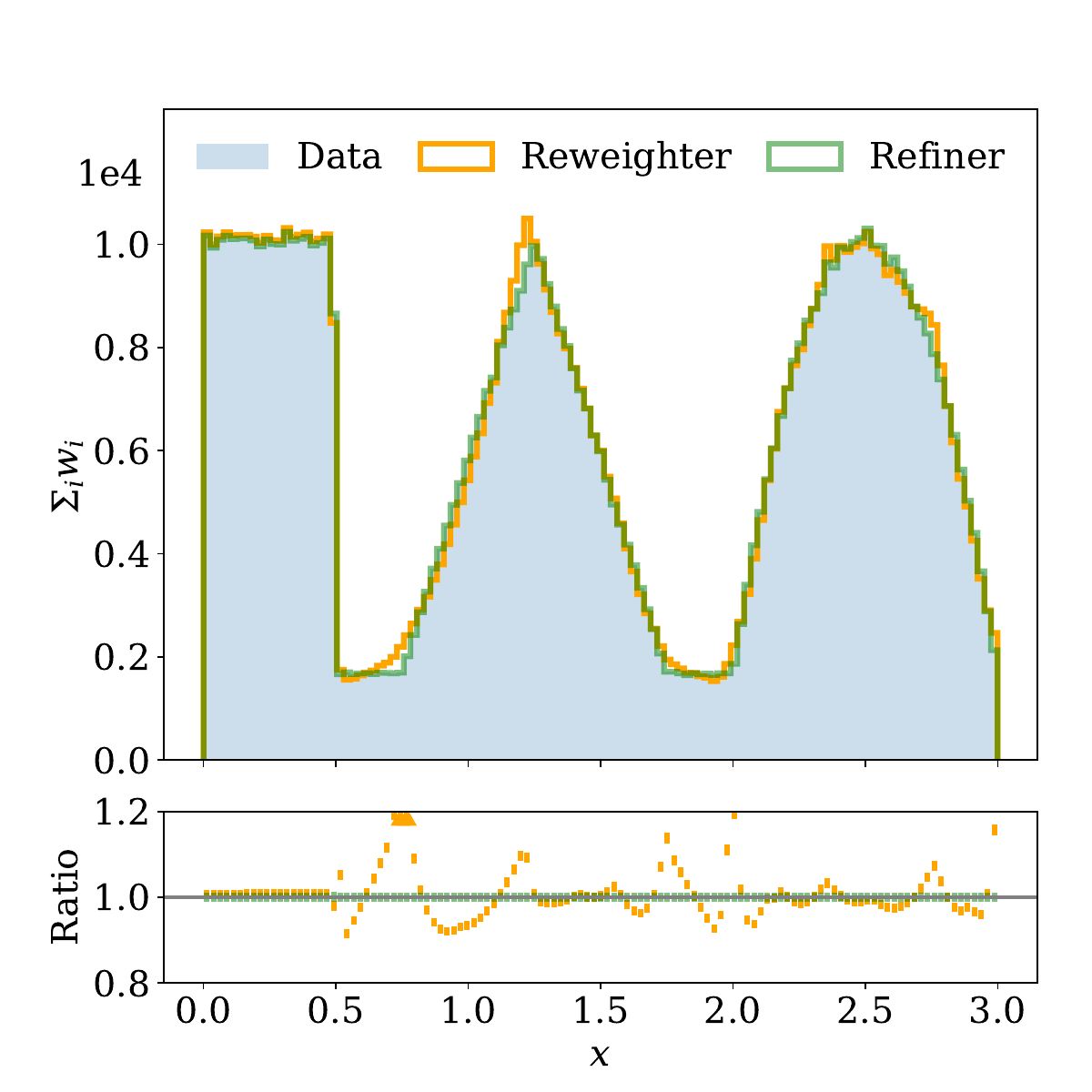}
  }%no space
  \hfill
  \subfloat[]{\label{fig:weight_shape:weights}
    \centering
    \includegraphics[width=0.32\linewidth]{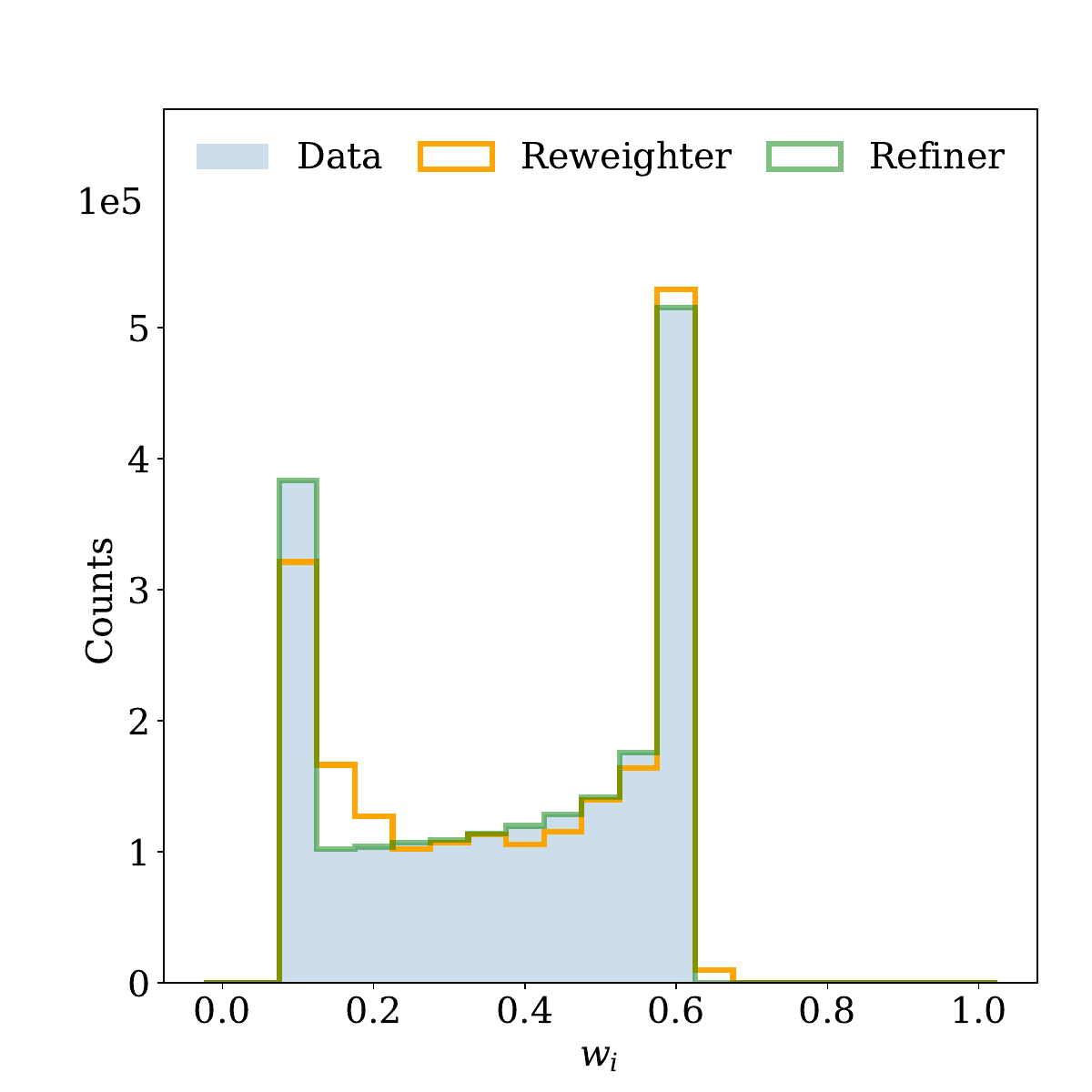}
  }
  \caption{
    Non-trivial, non-negative weight distribution.
    Left: a histogram of the synthetic observable broken down into positive and negative samples.
    Middle: the reweighted/refined spectrum compared to the initial data (with negative weights).
    Right: the distribution of sample weights before and after reweighting/refining. The error bars in the middle plot represent statistical uncertainties.
  }
  \label{fig:weight_shape}
\end{figure*}

This case study compares the performance of the refiner and the reweighter method on a case in which the sample weights follow a non-trivial shape.
For illustration, we consider a case where 10M samples are uniformly distributed.
Their weights follow a non-differentiable, positive distribution defined by a piecewise function comprising a rectangle, triangle, and inverted quadratic spline as shown in \Cref{fig:weight_shape:raw}.
Since all samples have positive weights, the distribution of positively weighted samples is identical to the overall sample distribution, and no weight transformation is required.
The refiner approach can exploit this, allowing the network to simply learn the identity function.
In contrast, the reweighter approach attempts to infer the entire weight distribution, necessitating the network to learn the non-trivial shape.

\Cref{fig:weight_shape:counts} and \cref{fig:weight_shape:weights} display the distributions of samples and weights before and after the weight transformations, respectively.
As anticipated, the refiner approach yields highly accurate sample and weight distributions, since it only requires learning the identity function.
In contrast, the sample and weight distributions of the reweighter approach exhibit various artifacts and mismodelings, resulting from the complex, non-trivial distribution of weights, which poses significant challenges for neural network modeling.
This study highlights the fact that the weight refinement can be less challenging and more accurate than corresponding methods, depending on the shape of the weights.

\subsection{Synthetic Spectrum Example}

This case study shows the performance of the refiner and the reweighter method on a case in which the weights have a stochastic spread in each particular phase space bin.
For illustration, the samples follow a distribution that can be described by two separate Gaussian functions.
Additionally, the sample weights within each phase space bin are also distributed according to two Gaussian functions.

\begin{figure*}
  \centering
  \subfloat[]{\label{fig:gauss_spread:raw}
    \centering
    \includegraphics[width=0.32\linewidth]{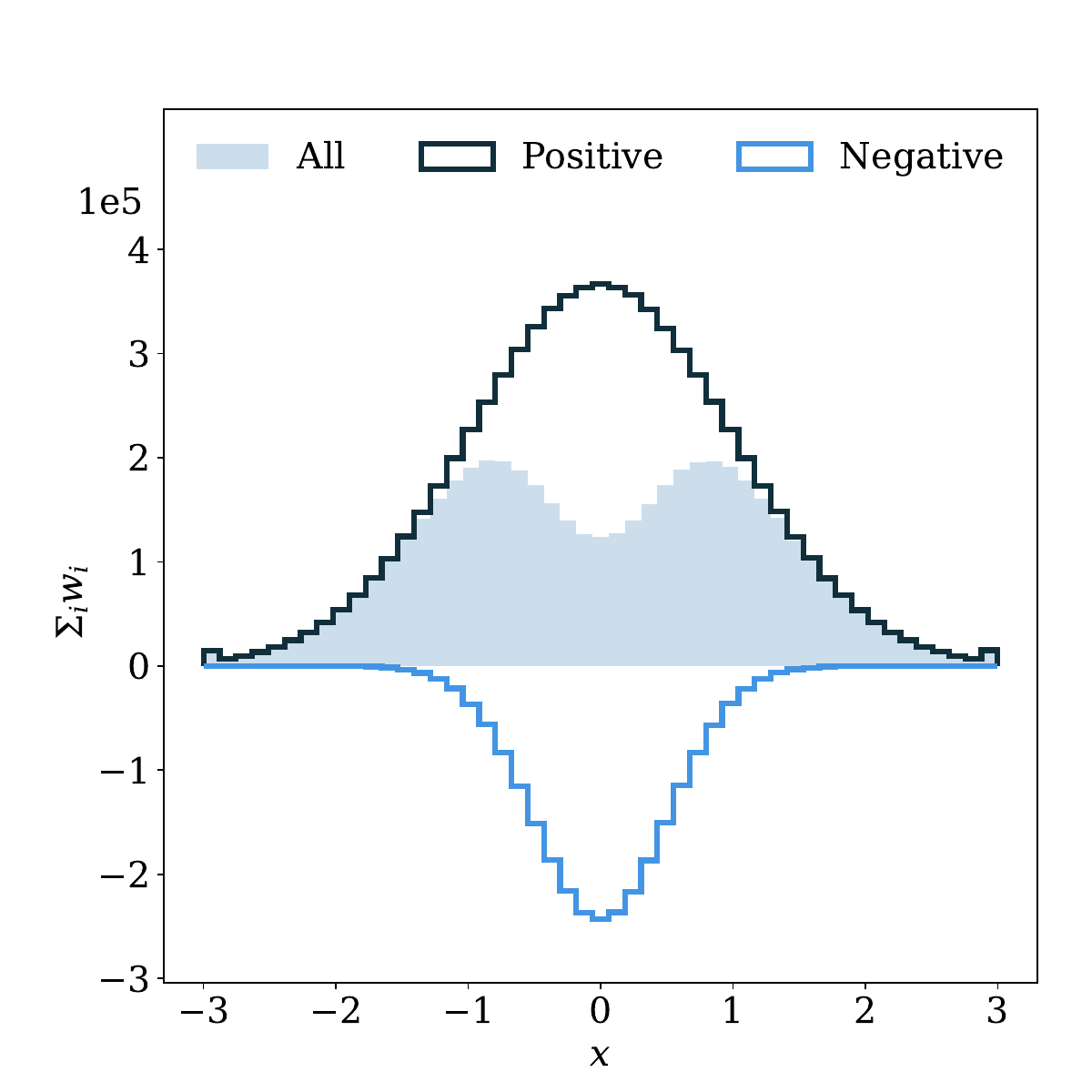}
  }%no space
  \hfill
  \subfloat[]{\label{fig:gauss_spread:counts}
    \centering
    \includegraphics[width=0.32\linewidth]{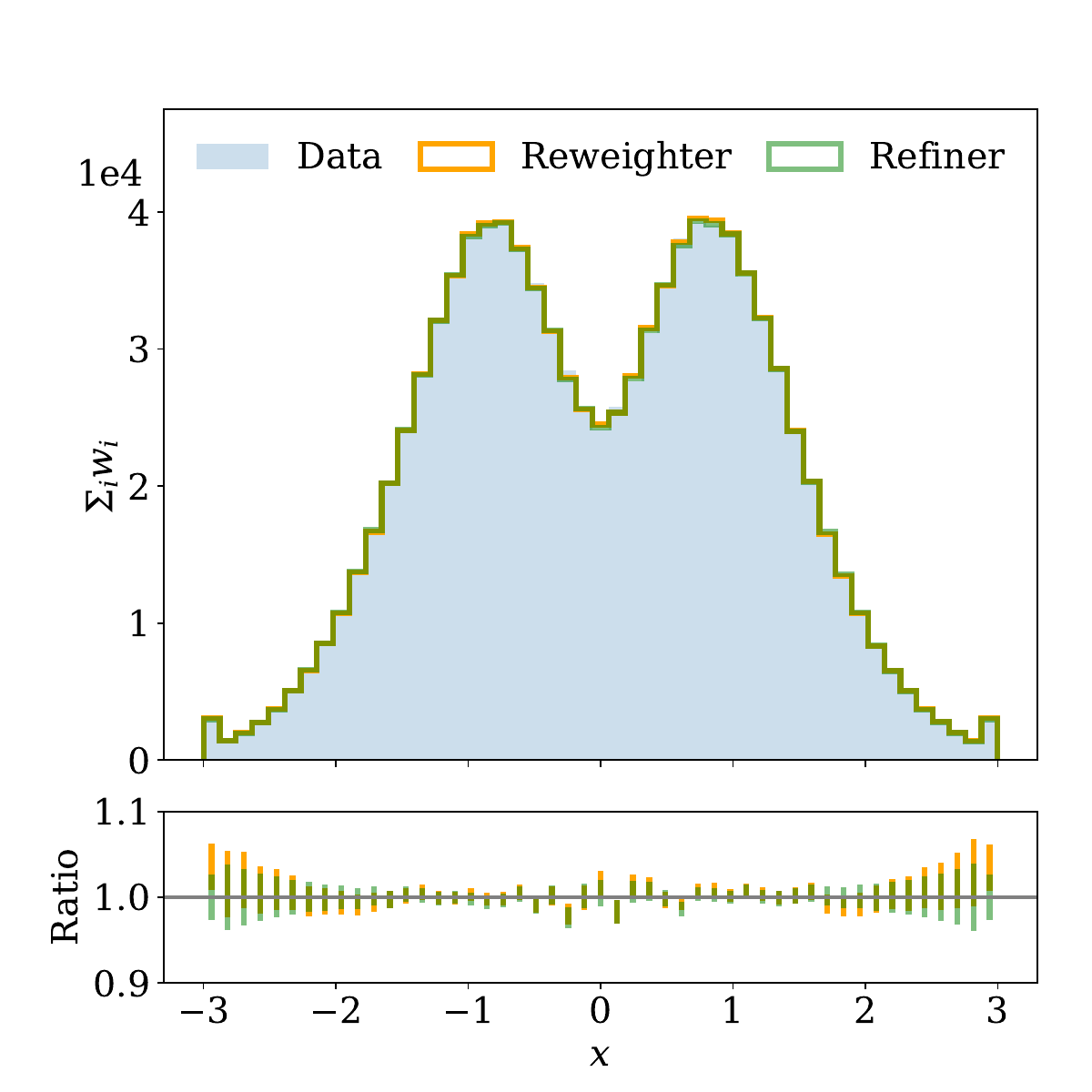}
  }%no space
  \hfill
  \subfloat[]{\label{fig:gauss_spread:scatter}
    \centering
    \includegraphics[width=0.32\linewidth]{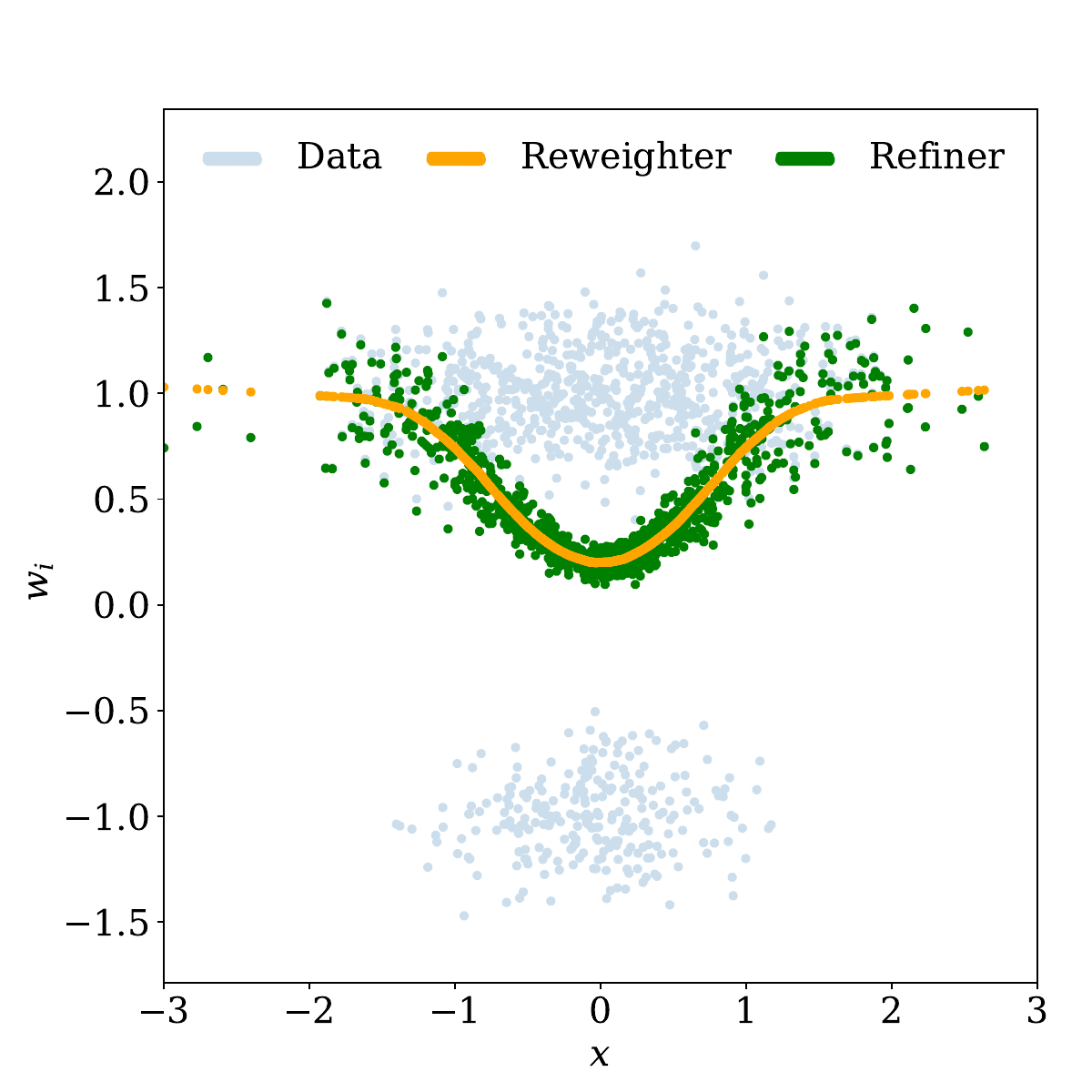}
  }
  \caption{
    Gaussian observable with Gaussian weights, including negative weights.
    Left: a histogram of the synthetic observable broken down into positive and negative samples.
    Middle: the reweighted/refined spectrum compared to the initial data (with negative weights).
    Right: the distribution of sample weights before and after reweighting/refining. The error bars in the middle plot represent statistical uncertainties.
  }
  \label{fig:gauss_spread}
\end{figure*}

In particular, the data consist of 7.5M samples following a Gaussian distribution centered at 0 with a width of 1 with weights following a Gaussian distribution centered at 1 with a spread of 0.2 and 2.5M samples following a Gaussian distribution centered at 0 with a width of 0.5 with weights following a Gaussian distribution centered at -1 with a spread of 0.2.
\Cref{fig:gauss_spread:raw} displays the distribution of samples with their original weights prior to the weight transformations, and the light-colored points in \Cref{fig:gauss_spread:scatter} show the corresponding weight distributions.

\Cref{fig:gauss_spread:counts} demonstrates that both transformations work and can restore the original distribution within the statistical uncertainties.
However, \Cref{fig:gauss_spread:scatter} shows that the reweighter approach predicts one weight for each phase space bin, while the refiner approach predicts a whole spectrum of weights in each phase space bin.
Doing this, the refiner approach preserves the relative spectrum of weights among the positive and pseudo-positive weights, respectively, in each phase space bin.
While not necessarily an advantage, it could be useful if downstream analysis makes use of the latent degrees of freedom that produce the stochastic weight distribution.

\subsection{Synthetic Extrapolation Example}

This case study demonstrates the extrapolation capabilities of the reweighter and refiner methods to extend beyond the region where the majority of samples are concentrated and the weight distribution can be readily learned.
For illustration, we use the same dataset as in the previous section, but with no spread in weights for the positive and negative weight samples.
This allows us to focus on the tails of the distribution to explore the extrapolation properties.

\begin{figure*}
  \centering
  \subfloat[]{\label{fig:gauss_easy:raw}
    \centering
    \includegraphics[width=0.32\linewidth]{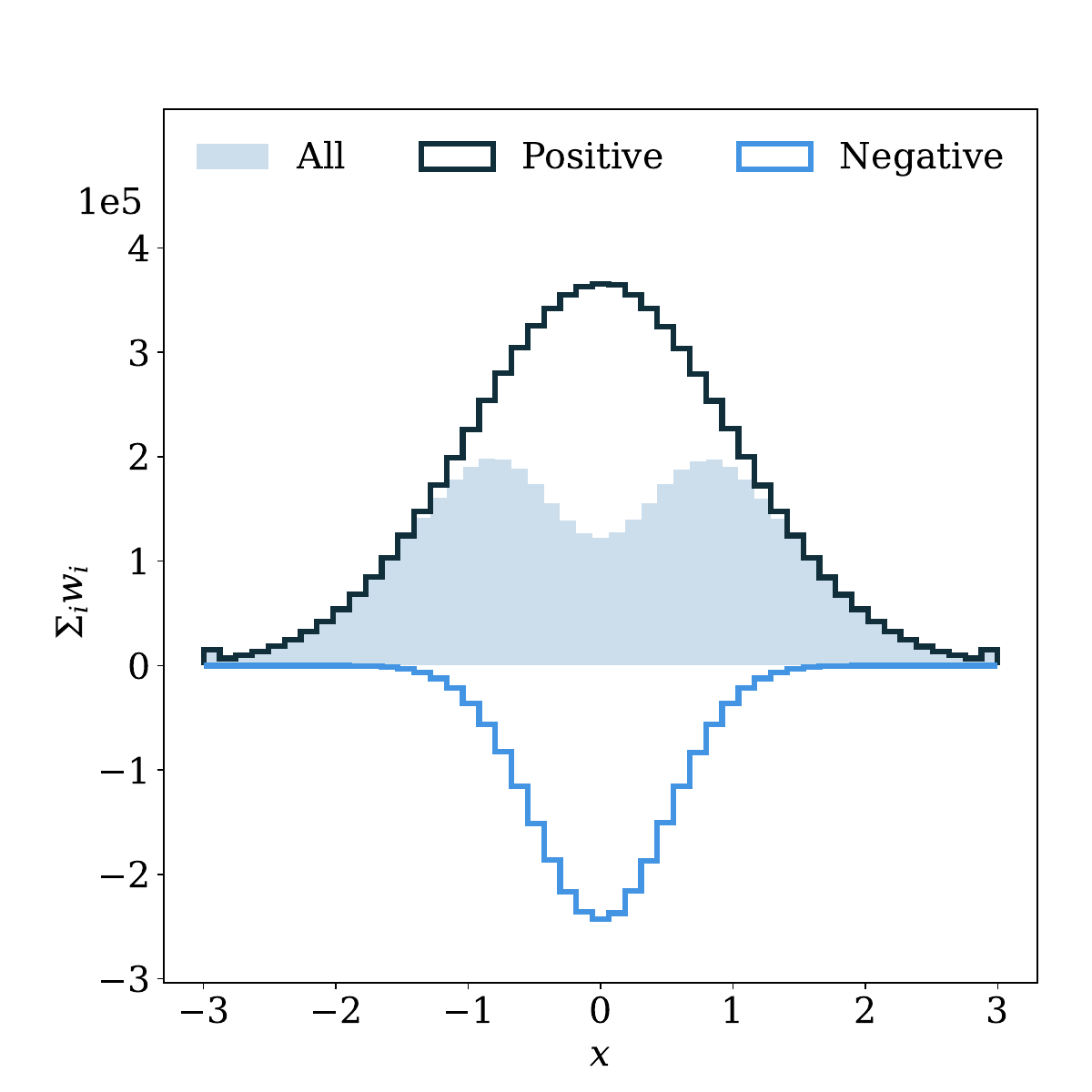}
  }%no space
  \hfill
  \subfloat[]{\label{fig:gauss_easy:counts}
    \centering
    \includegraphics[width=0.32\linewidth]{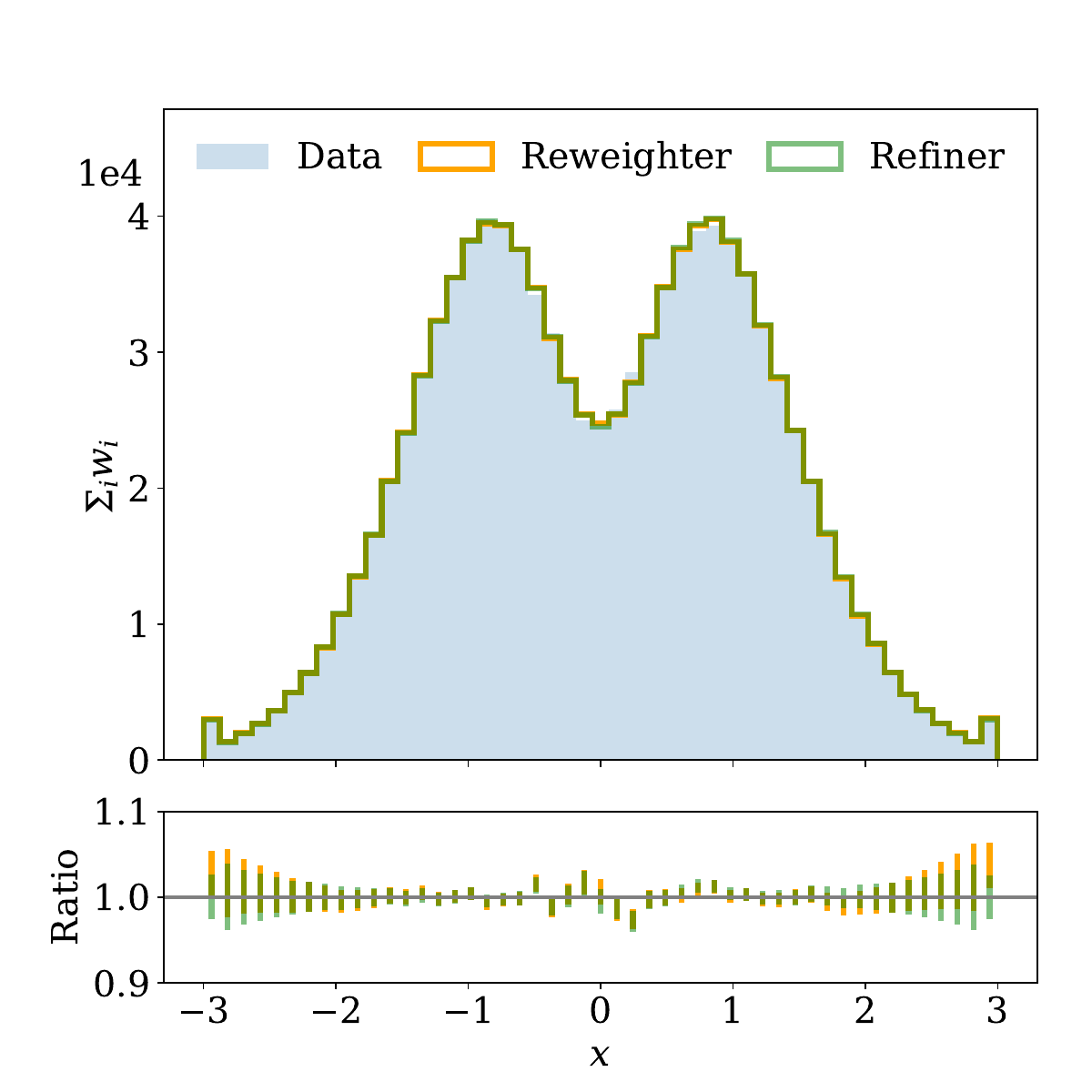}
  }%no space
  \hfill
  \subfloat[]{\label{fig:gauss_easy:scatter}
    \centering
    \includegraphics[width=0.32\linewidth]{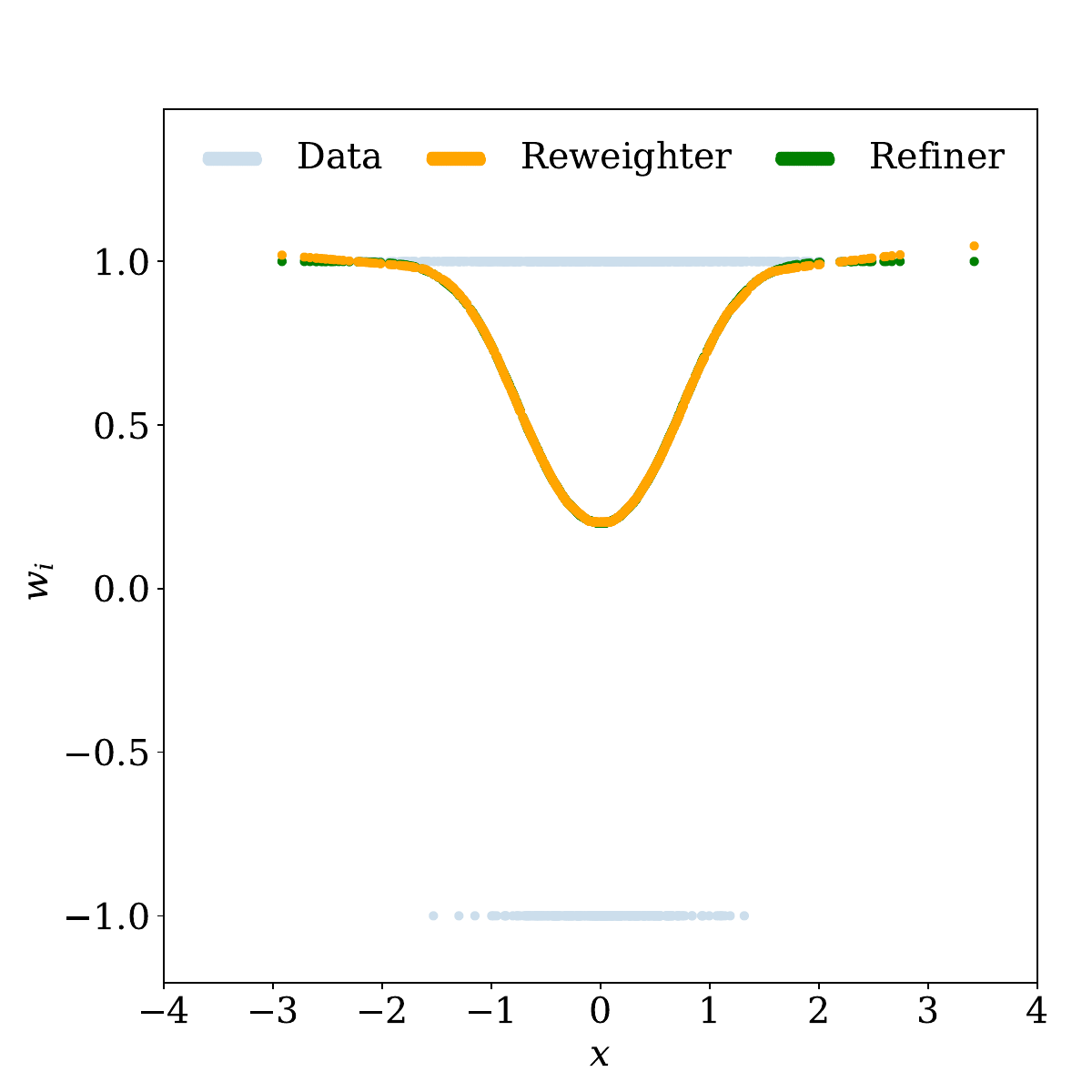}
  }
  \caption{
    Gaussian observable with non-random weights, including negative weights.
    Left: a histogram of the synthetic observable broken down into positive and negative samples.
    Middle: the reweighted/refined spectrum compared to the initial data (with negative weights).
    Right: the distribution of sample weights before and after reweighting/refining. The error bars in the middle plot represent statistical uncertainties.
  }
  \label{fig:gauss_easy}
\end{figure*}

In particular, the data consist of 7.5M samples following a Gaussian distribution centered at 0 with a width of 1, with weights set to 1, and 2.5M samples following a Gaussian distribution centered at 0 with a width of 0.5, with weights set to -1.
\Cref{fig:gauss_easy:raw} displays the distribution of samples with their original weights prior to the weight transformations.

\Cref{fig:gauss_easy:counts} displays the sample distributions before and after weight transformations.
Both transformations are effective in restoring the original distribution within statistical uncertainties.
However, a systematic up-fluctuation is evident on the sides of the distribution for the reweighter approach.
This corresponds to a weight function in \cref{fig:gauss_easy:counts} that increases in both directions away from zero, when it should instead be a constant.
This case study demonstrates that the refiner approach can have better extrapolation capabilities than corresponding methods, depending on the shape of the weights.

\subsection{Synthetic Negative Density Example}

This case study compares the performance of the reweighter and the refiner method on a case in which some part of the phase space has a negative density.
While this is not possible for physical cross sections, scenarios with negative sample density can arise in various situations.
For instance, limited statistics can lead to negative densities in certain regions of phase space, even if the underlying distribution is non-negative.
Additionally, certain analysis techniques may involve non-physical processes with negative cross sections, such as events representing interference terms.
Furthermore, the breakdown of perturbation theory near soft/collinear singularities can result in negative cross sections.

Some of these problems can be resolved with separate techniques, such as addressing issues related to perturbation theory breakdown through methods like matching to a parton shower, or by equipping the reweighter with a more sophisticated loss function that explicitly handles regions of negative density \cite{drnevich2024neuralquasiprobabilisticlikelihoodratio}.
In contrast, the refinement approach inherently handles regions with negative weights without requiring additional modifications or mitigations, offering a straightforward solution to these challenges.

It is not possible to render all weights non-negative in a phase space region with an overall negative underlying density.
However, it should still be possible to apply a weight transformation method to the overall phase space, allowing for a graceful transformation of weights in the remaining phase space.
As demonstrated below, the refiner method exhibits this desirable property, in contrast to the reweighter approach.

\begin{figure*}
  \centering
  \subfloat[]{\label{fig:gauss_negative:raw}
    \centering
    \includegraphics[width=0.32\linewidth]{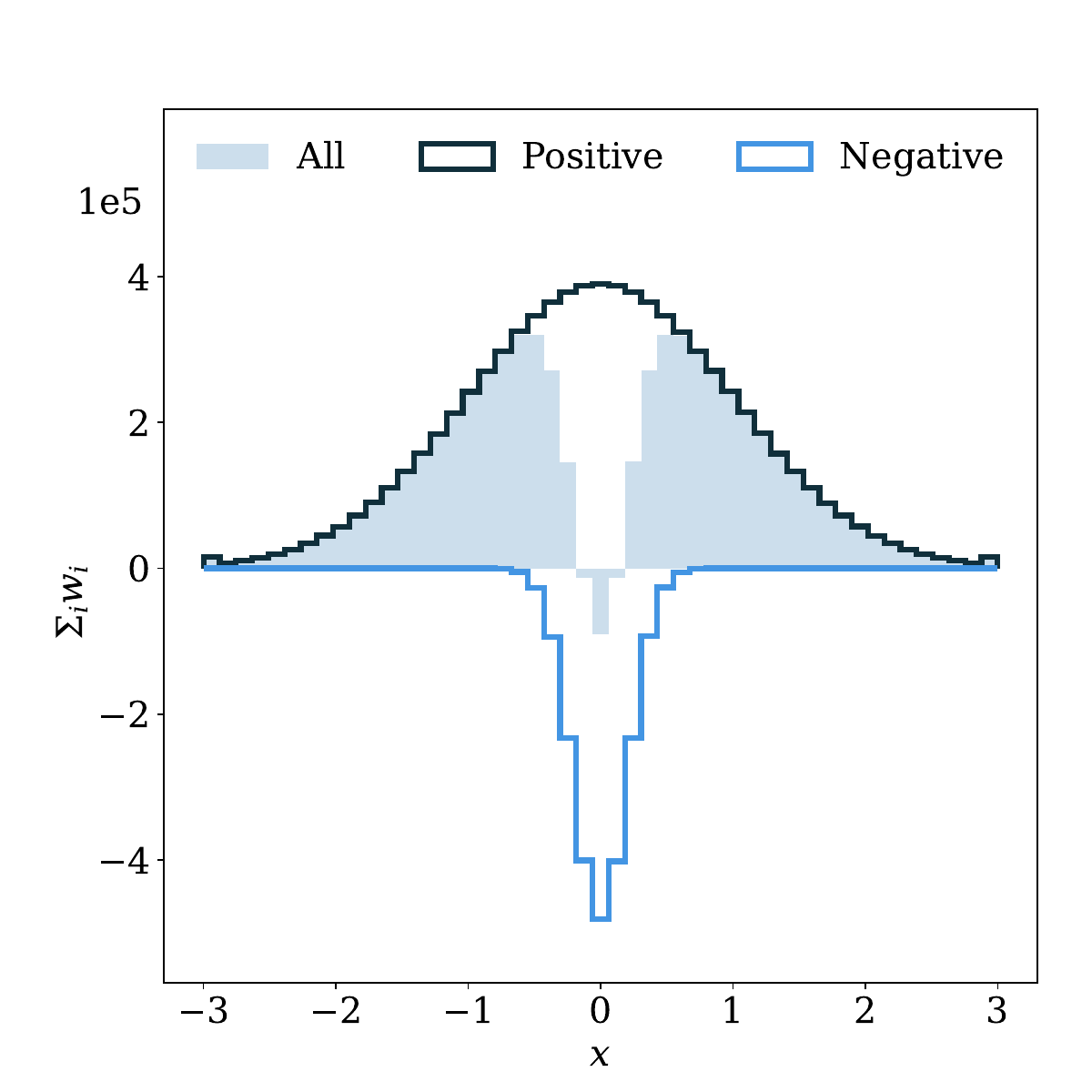}
  }%no space
  \hfill
  \subfloat[]{\label{fig:gauss_negative:counts}
    \centering
    \includegraphics[width=0.32\linewidth]{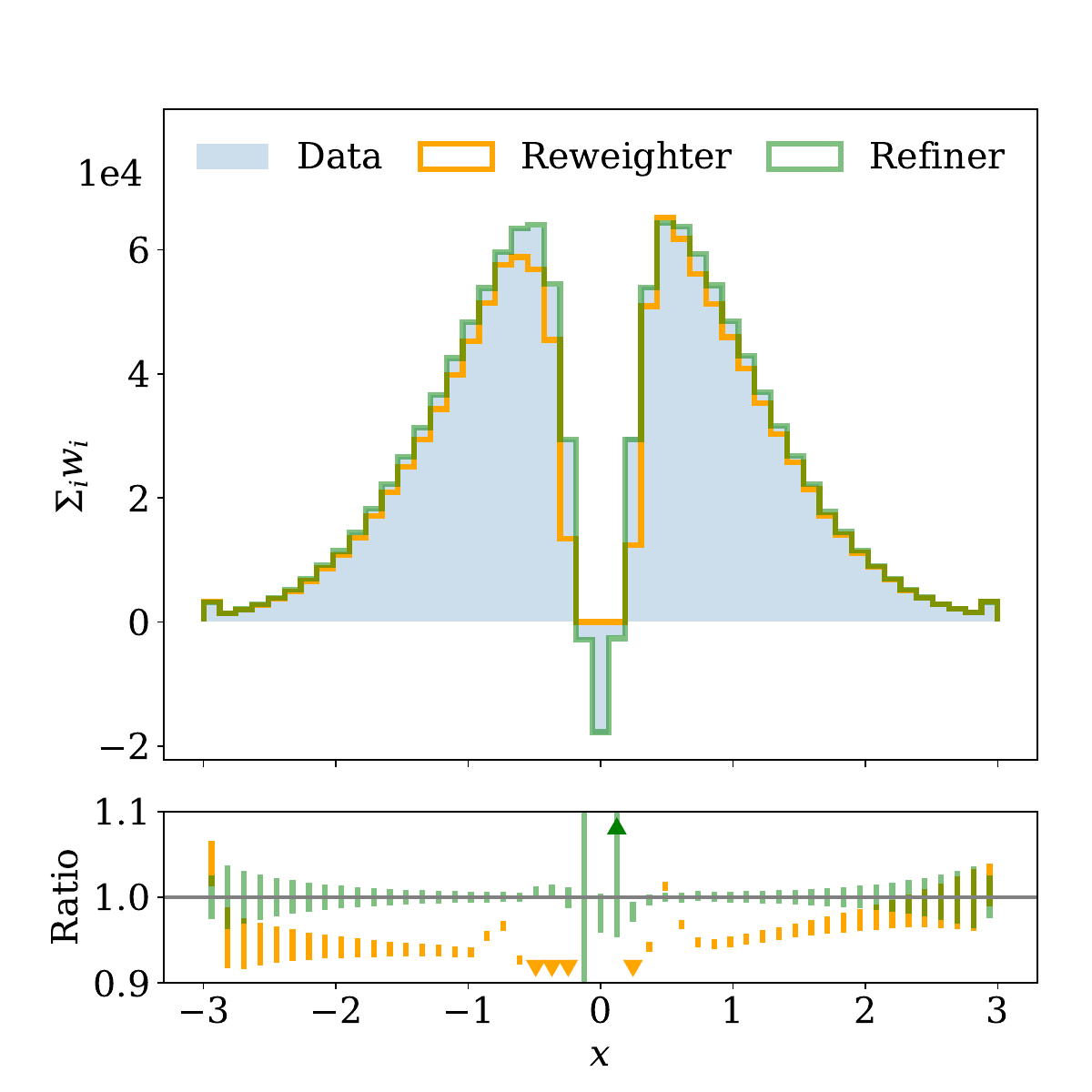}
  }%no space
  \hfill
  \subfloat[]{\label{fig:gauss_negative:training}
    \centering
    \includegraphics[width=0.32\linewidth]{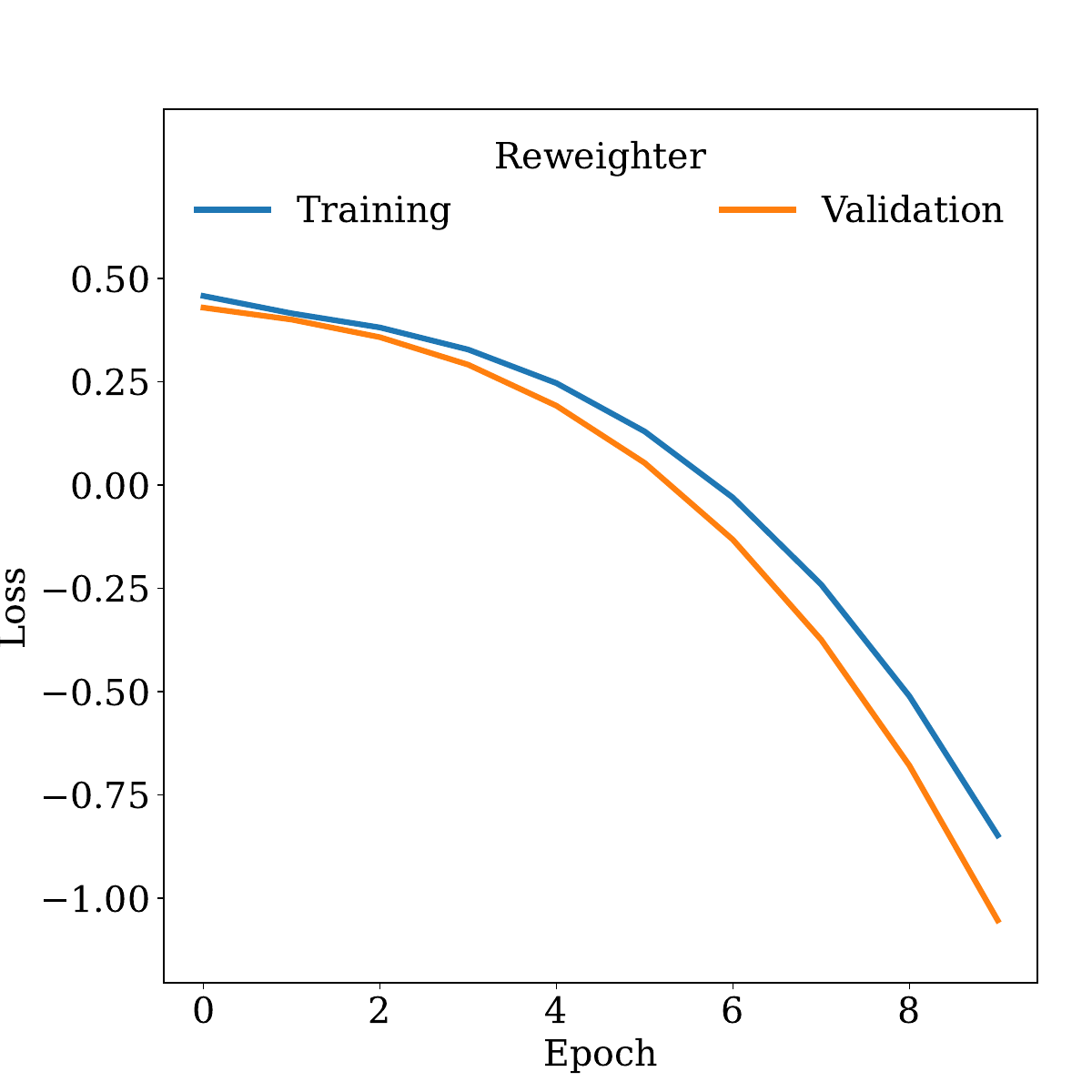}
  }
  \caption{
    Gaussian observable with regions of negative cross section.
    Left: a histogram of the synthetic observable broken down into positive and negative samples.
    Middle: the reweighted/refined spectrum compared to the initial data (with negative weights).
    Right: the distribution of sample weights before and after reweighting/refining. The error bars in the middle plot represent statistical uncertainties.
  }
  \label{fig:gauss_negative}
\end{figure*}

For illustration, we generate 7.5M samples following a Gaussian distribution centered at 0 with a width of 1, with weights set to 1, and 2M samples following a Gaussian distribution centered at 0 with a width of 0.2, with weights set to -1.
\Cref{fig:gauss_negative:raw} displays the distribution of samples with their original weights prior to the weight transformations.
It can be seen that the density of the weighted samples is negative in the range from $-0.14$ to $0.14$.

\Cref{fig:gauss_negative:counts} shows the sample distributions before and after the weight transformations.
To illustrate the divergent behavior of the reweighter approach, its neural network is deliberately trained using the Adam optimizer with a fixed learning rate of $2\times10^{-5}$, not hiding its divergent behavior with a decaying learning rate.
The refiner approach can accurately model the entire distribution, including the region of phase space with negative sample density.
In contrast, the reweighter approach, by design, is constrained to produce non-negative weights and therefore fails to capture the distribution in the region of negative density.

Although one might argue that the reweighter's inability to model negative densities is a desirable feature, the method ultimately fails to adequately capture the remaining part of the phase space.
This is because, in the presence of a phase space with negative density, the reweighter becomes overly focused on this region, causing the training loss to diverge negatively, as illustrated in \cref{fig:gauss_negative:training}.
Even a small fraction of negative phase space can be sufficient to compromise the entire reweighter method.
If the network is able to identify this problematic region, the training process will diverge, rendering the reweighter approach unusable. 
In contrast, the refiner can assign negative weights to samples in these regions of phase space and still work properly everywhere else.

This case study demonstrates the resilience of the refiner method in handling phase space regions with negative density, compared to the corresponding reweighter method, which does not have this feature.

\section{Sampling Uncertainties}
\label{sec:resampling}

Section~\ref{sec:methods} showed how to use reweighting (apostrophe) or refining (tilde) so that the local weight $\langle W|X=x\rangle=\langle W'|X=x\rangle=\langle \tilde{W}|X=x\rangle$, where capital letters are random variables and lower-case variables are realizations of those random variables.
With a resampling protocol, it is also possible to match the statistical uncertainty, in the form of the second moment: $\langle W^2|X=x\rangle$.
Ref.~\cite{Nachman:2020fff} proposed to accomplish this by randomly discarding samples with probability $\theta(x)$ and then rescaling the weight at $X=x$ by $k(x)$, $w\mapsto w\,k(x)$.
If $\theta(x)$ is a Bernoulli random variable with rate $\lambda(x)$, we can compute:

\begin{align}
    \langle k(X)& \theta(X) \hat{W} | X=x \rangle\\
    &= k(x) \langle \theta(X) | X=x \rangle \langle \hat{W} | X=x \rangle \\
    &= k(x) \lambda(x) \langle \hat{W} | X=x \rangle \,,
\end{align}
where $\hat{W}$ is $W'$ or $\tilde{W}$.
For the resampling to preserve the mean, $\lambda(x)=1/k(x)$.
We can now compute the second moment to determine $k(x)$:

\begin{align}
    \langle (k(X)& \theta(X) \hat{W})^2 | X=x \rangle\\
    &= k(x)^2 \langle \theta(X)^2 | X=x \rangle \langle \hat{W}^2|X=x \rangle \\
    &= k(x)^2 \langle \theta(X) | X=x \rangle \langle \hat{W}^2|X=x \rangle \\
    &= k(x)^2 \lambda(x) \langle \hat{W}^2 | X=x \rangle\\
    &= k(x) \langle \hat{W}^2 | X=x \rangle\,,
\end{align}
which implies that $k(x)=\langle W^2|X=x\rangle/\langle \hat{W}^2|X=x\rangle$.
For reweighting, $\hat{W}=\langle W|X=x\rangle$, so 

\begin{align}
    k_\text{reweight}(x)=\frac{\langle W^2|X=x\rangle}{\langle W|X=x\rangle^2}\,.
\end{align}
This requires estimating the local second moment in addition to the local first moment.
The denominator of $k$ for the analogous refining formula further requires estimating the local second moment of $W$, since $W$ is not constant given $X=x$.
This is possible to achieve in practice, but requires significant overhead.

In this section, we propose a new approach to resampling by promoting $k$ to be a random variable $K$ that depends on both $W$ and $X$.
In particular, we define 

\begin{align}
\label{eq:newk}
    K(W,X)=\frac{W^2}{\hat{W}^2}\,,
\end{align}
where the $X$ dependence is inside $\hat{W}$. Then:

\begin{align}
    \langle (K & \theta(K) \hat{W} )^2 | X=x \rangle\\
    &=\langle W^4 \hat{W}^{-4} \theta(K)^2 \hat{W}^2 | X=x \rangle\\
    &=\langle W^4 \hat{W}^{-2} \theta(K) | X=x \rangle\\
    &=\langle \langle W^4 \hat{W}^{-2} \theta(K) | W=w, X=x \rangle | X=x \rangle\\
    &=\langle \lambda(K) \langle W^4 \hat{W}^{-2} | W=w, X=x \rangle | X=x \rangle\\
    &=\langle \hat{W}^{2} W^{-2} W^4 \hat{W}^{-2} | X=x \rangle\\
    &=\langle W^2|X=x\rangle\,,
\end{align}
where the double expectation value is a version of the law of total probability.
In contrast to the previous approach, the new resampling defined by Eq.~\ref{eq:newk} does not require estimating any new quantities.

\begin{figure*}
  \centering
  \subfloat[]{\label{fig:gauss_spread_resampled:variances}
    \centering
    \includegraphics[width=0.32\linewidth]{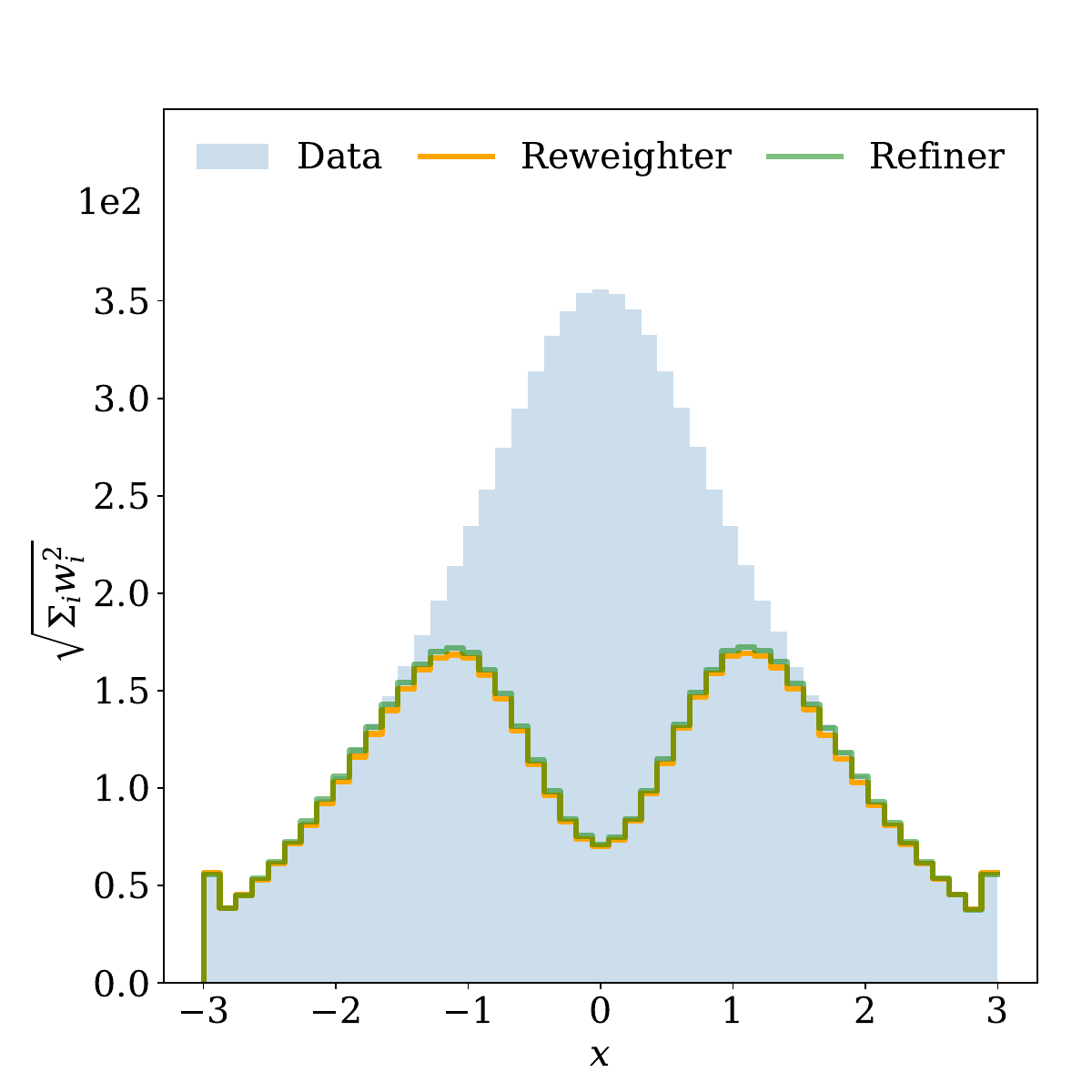}
  }%no space
  \hfill
  \subfloat[]{\label{fig:gauss_spread_resampled:counts_resampled}
    \centering
    \includegraphics[width=0.32\linewidth]{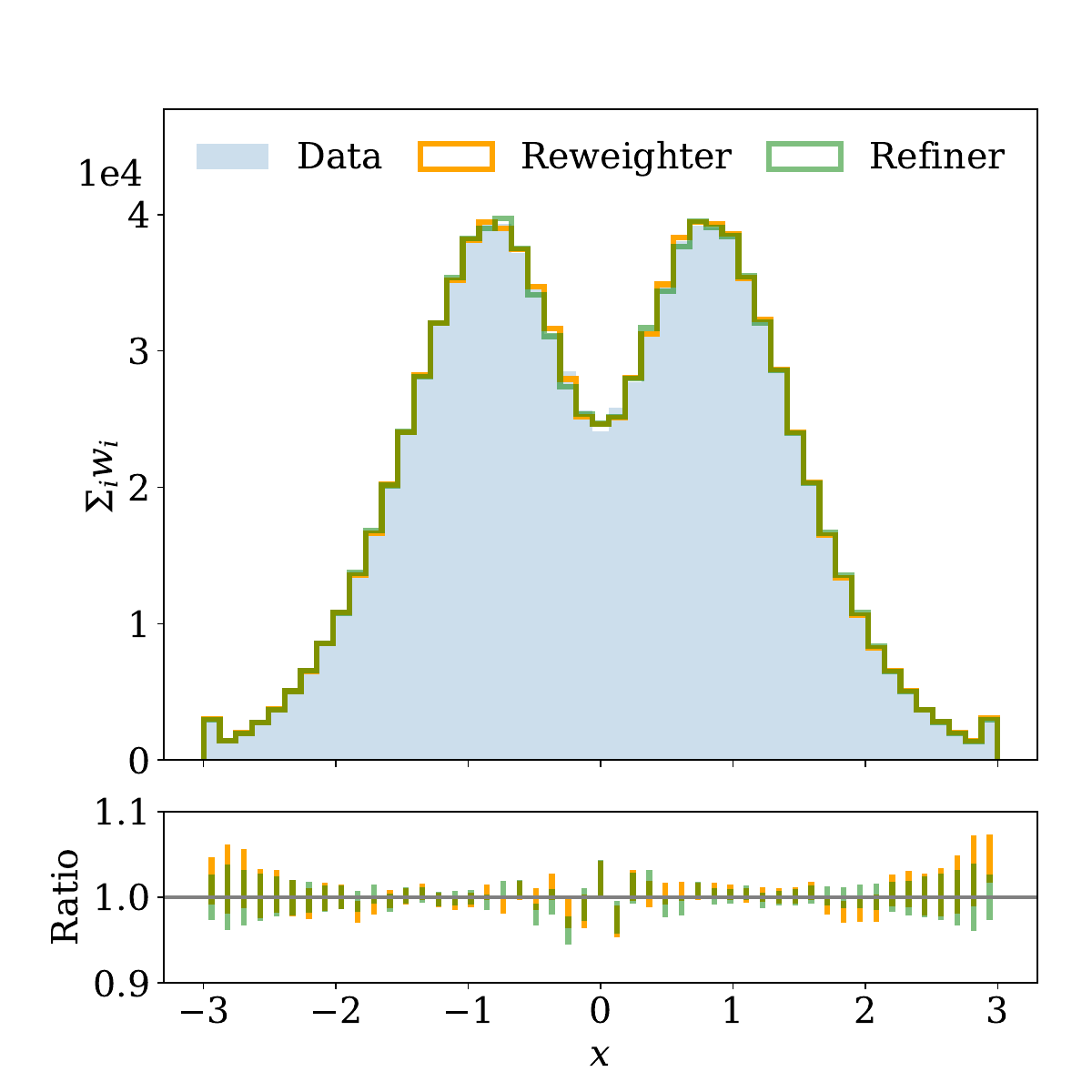}
  }%no space
  \hfill
  \subfloat[]{\label{fig:gauss_spread_resampled:variances_resampled}
    \centering
    \includegraphics[width=0.32\linewidth]{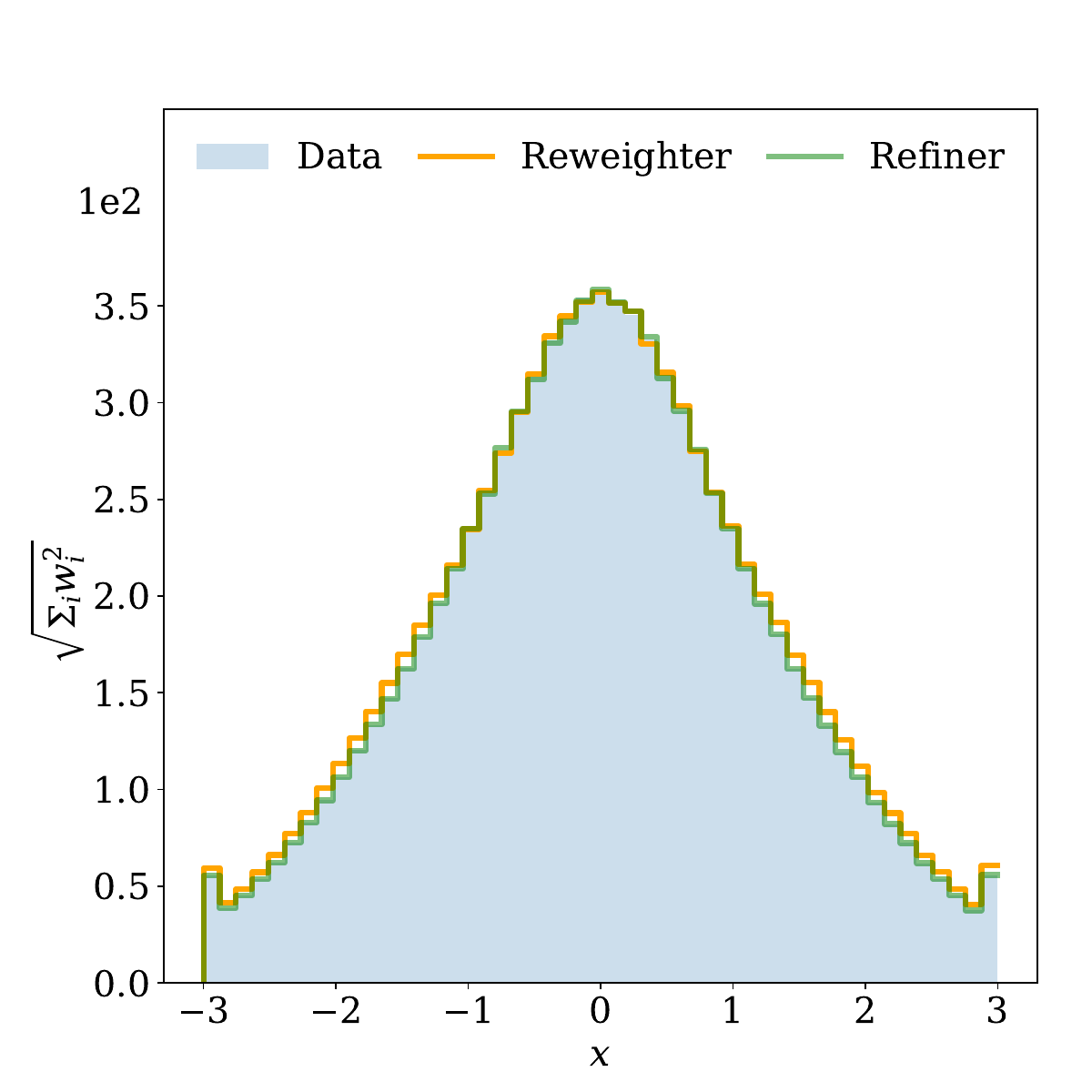}
  }
  \caption{
    Gaussian observable with Gaussian weights, including negative weights.
    Left: a histogram representing the uncertainties of the reweighted/refined spectrum without the resampling compared to the initial data.
    Middle: the reweighted/refined spectrum with the resampling compared to the initial data (with negative weights).
    Right: a histogram representing the uncertainties of the reweighted/refined spectrum with the resampling compared to the initial data.
  }
  \label{fig:gauss_spread_resampled}
\end{figure*}

\Cref{fig:gauss_spread_resampled} illustrates the operation of the new resampling method applied to case study C.
\Cref{fig:gauss_spread_resampled:variances} shows the standard Monte Carlo estimation of uncertainties, derived from the square root of the sum of squared sample weights, after applying the reweighting or refining method without any resampling.
It can be seen that the uncertainties are inaccurately represented after simply applying the reweighting or refining method.
\Cref{fig:gauss_spread_resampled:counts_resampled} and \cref{fig:gauss_spread_resampled:variances_resampled} display the sample distribution and uncertainty estimate after applying the reweighting or refining method with the resampling.
These figures indicate that the resampled distributions still match the original spectrum and now correctly represent the uncertainties.
Similar outcomes are observed for the other case studies.

\section{Conclusion and Outlook}
\label{sec:conclusions}

This paper introduces a novel method for transforming sample weights in a weighted dataset, with the goal of rendering them all positive, and a new resampling protocol to restore the statistical uncertainties of the initial samples.

Our new refinement approach scales the original weights of each sample using a phase space scaling transformation, offering a simpler learning task compared to previous methods that model the full weight distribution.
The paper introduces the method and demonstrates its advantages and comparable or superior performance on multiple physical and synthetic datasets.
These advantages include a simplified learning task, preservation of the spectrum of positive and negative weights in each phase space bin, improved extrapolation, and robustness to phase space regions with negative density.
While the improvements on current datasets are modest, they might be particularly relevant in the context of increasingly stringent demands on accuracy and precision of simulations in future experiments.

Our new resampling protocol restores the statistical uncertainties of the distribution by randomly discarding samples with a probability that depends on the initial sample weight and phase space, followed by a rescaling of the weights.
In contrast to previous approaches, this method avoids estimating the local second moment of the weights, simplifying implementation and reducing computational cost.
The resampling protocol is a straightforward replacement of existing methods and is compatible with many different weight transformations.

Even though we have focused on cross sections, the refining methodology may be useful more generally when datasets have negative weights.
Beyond the task of transforming sample weights, this work also illustrates the importance of considering alternative strategies for likelihood-free inference even when all strategies have the same formal, asymptotic behavior. 

\section*{Code Availability}

The code for this project is available at: \url{https://github.com/Nollde/weight_refiner}.
The top quark datasets used in this study are available upon request to the authors.

\section*{Acknowledgments}
We would like to thank Jesse Thaler for valuable comments on the manuscript.
This work is supported by the U.S. Department of Energy (DOE), Office of Science under contract DE-AC02-05CH11231. 

\bibliography{apssamp}
\bibliographystyle{JHEP}

\end{document}